\documentclass[10pt]{iopart}
\usepackage{iopams,bm,cite,color,graphicx,mathptmx,url,hyperref}

\newcommand{\half}{\mbox{$\textstyle \frac{1}{2}$}}
\newcommand{\ket}[1]{\left| #1 \right\rangle}
\newcommand{\bra}[1]{\left\langle #1 \right|}
\newcommand{\proj}[1]{\ket{#1}\bra{#1}}

\eqnobysec

\begin{document}

\title[Correlations in emitters coupled to plasmonic waveguides]{Correlations in emitters coupled to plasmonic waveguides}

\author{C~E~Susa$^{1\dagger}$, J~H~Reina$^1$ and
  L~L~S\'anchez-Soto$^2$}

\address{$^1$~Departamento de F\'isica, Universidad del Valle,
 A.~A. 25360, Cali, Colombia}

\address{$^2$~Departamento de \'Optica, Facultad de F\'{\i}sica,
  Universidad Complutense, 28040~Madrid, Spain}

\ead{$^\dagger$cristian.susa.q@correounivalle.ed.co}

\begin{abstract}
  We report on quantum, classical, and total correlations in a set of
  distant quantum emitters coupled via their interaction with the
  plasmon modes of a one-dimensional waveguide driven by an external
  laser field.   The coupling of the emitters with the plasmonic modes and its
  influence on the collective decay rate suggest that entanglement does
  not play a significant role in the qubit dynamics. Rather, discord
  is the quantity that matters and should be harnessed as a
  resource.

\end{abstract}



\section{Introduction}
Entanglement is a cornerstone of quantum
theory~\cite{horodecki,martini,lukin,polzik}.  It has strong
practical implications for futuristic quantum
technologies~\cite{obrien,bennett,knill, bennett1,gisin, man} and,
consequently, it has been investigated in a wide variety of physical
systems~\cite{ladd,Schoelkopf,jhQC,makhlin}.  For a thorough
description of this phenomenon, a formal framework has been introduced
over the past decade to quantify quantum~\cite{Zurek} and
classical~\cite{Vedral} correlations in a physical system. The former,
being of a more general character than entanglement, has also been
hinted to play a role in quantum information processing through 
the so-called mixed state-based quantum computing~\cite{dqc1,datta}.
 
Quantum emitters (e.g., single molecules~\cite{Hettich,jh3} or quantum
dots~\cite{soren1,jhdotspin,jh4}) coupled to surface plasmons of conducting
nanowires or waveguides have been seen as a promising hardware for
quantum information processing because of the strong coupling that can
be achieved between the emitters and the plasmons \cite{soren1}. The
collective spontaneous decay and the plasmon-mediated emitter-emitter
coupling have been investigated for a two-qubit (emitter) system close
to a nanowire~\cite{soren2,david}, as has the resonance energy
transfer mediated by different plasmonic nanowaveguides~\cite{nano}.
Entanglement of two qubits, mediated by a V-groove plasmonic
waveguide, has recently been theoretically
studied~\cite{tudela,prb11}  and quantified via the
concurrence~\cite{wootter}. We address this problem within a more
general framework: we calculate the entire spectrum for the
entanglement, classical, quantum, and total
correlations~\cite{popescu,hamieh,Vedral2} for emitters coupled
to plasmonic modes.

We demonstrate that it is not entanglement but the quantum
correlations (measured by the discord) that provide the relevant
robust features that could be harnessed for plasmon-assisted
information processing. Our results support the conclusion that
the entanglement of formation~\cite{wootter} is a natural metric
  and that, in the context of entropic measures, quantification via
  the concurrence might be inappropriate.  

\section{Quantum emitters, plasmon modes and correlations}
\label{system}

We start by briefly delineating the model. We are dealing with two
two-level atoms fixed at positions $\mathbf{r}_{i}$ ($i = 1,2$), with
transition frequencies $\omega_{i}$ and separated by the vector
$\mathbf{r}_{12}$. We denote by $\ket{0_{i}}$ and  $\ket{1_{i}}$ the
ground and excited states of the emitter $i$, with associated
transition dipole moments $ \hat{\mathbf{\mu}}_{i} \equiv \bra{0_i} 
\mathbf{D}_{i} \ket{1_i}$,  $\mathbf{D}_{i}$ being the corresponding
dipole operators. 

The emitters are embedded in a medium of refractive index $n$ and
interact via a dipole-dipole coupling, so that the system Hamiltonian 
$H_{S}$ can be written as
\begin{equation}
  \label{eq:1}
  {H}_{S} = {H}_{0} + {H}_{12} \, ,
\end{equation}
where the free and interaction Hamiltonians are
\begin{eqnarray}
  {H}_{0} & = & \half \hbar \omega_{1} \sigma_{z}^{(1)} + 
 \half \hbar  \omega_{2} \sigma_{z}^{(1)}  \nonumber \\
 &  & \\
  {H}_{12} & = & \half \hbar V \left ( 
    \sigma^{(1)}_{x}\otimes\sigma^{(2)}_{x} 
    + \sigma^{(1)}_{y}\otimes\sigma^{(2)}_{y} 
  \right ) \, . \nonumber 
\end{eqnarray} 
The strength $V$ depends of the configuration of the interacting
dipoles and $\bm{\sigma}^{(i)}$ are the Pauli operators.

In addition, an external laser field of frequency $\omega_{L}$ drives the emitters, which we
represent by
\begin{eqnarray}
  \label{eq:2}
  H_{L} & = & \hbar \ell^{(1)}  \left ( \sigma_{-}^{(1)} e^{i \omega_{L} t} +
    \sigma_{+}^{(1)} e^{- i \omega_{L} t} \right ) 
     +   \hbar \ell^{(2)}  \left ( \sigma_{-}^{(2)} e^{i \omega_{L} t} +
    \sigma_{+}^{(2)} e^{- i \omega_{L} t} \right ) \, .
\end{eqnarray}
Here $\ell^{(i)}$ is the strength of this coupling, and is given by
$\hbar \ell^{(i)} = - \hat{\mu}_{i} \cdot \mathbf{E}_{i}$, with
$\mathbf{E}_{i}$ being the amplitude of the coherent field acting on
the $i$th emitter, and $\sigma^{(i)}_{+}=\ket{1_{i}}\bra{0_{i}}$, and
$\sigma^{(i)}_{-}=\ket{0_{i}}\bra{1_{i}}$ are the raising and lowering
Pauli operators acting on the $i$th emitter.

\subsection{Collective decay effects}

We next assume that the emitters are coupled with a bath of the
radiation field, so that the corresponding dissipative dynamics is
given by the total Hamiltonian ${H}={H}_S + {H}_{L}$, by means of the
quantum master equation~\cite{jh3}
\begin{eqnarray}
  \label{master}
  {\dot\rho} & = &  -\frac{i}{\hbar} [ {H}, {\rho} ] \nonumber \\
  &  - &   \sum_{i,j=1}^{2}\frac{\Gamma _{ij}}{2} \left( 
    {\rho} \sigma^{(i)}_{+} \sigma^{(j)}_{-} + 
    \sigma^{(i)}_{+}\sigma^{(j)}_{-}{\rho}
    -2\sigma^{(i)}_{-}{\rho} \sigma^{(j)}_{+} 
  \right)  ,
\end{eqnarray}
Without loss of generality, we set $\Gamma_{ii}\equiv \Gamma_{i}=\Gamma$, and
$\Gamma_{ij}=\Gamma^{\ast}_{ji}\equiv \gamma$  ($i\neq j$), the
individual, and collective spontaneous emission rates, respectively.

The explicit form of the individual emitter decay rate
$\Gamma_{i}$, as well as the collective decay rate $\gamma$ associated
to the dipole-dipole (qubit-qubit) interaction $V$ depend on the
particular physical setup under consideration.  For the case of 
 interacting `bare' quantum emitters, we have 
\begin{eqnarray}
  V & = &
  \label{V}
  \frac{3\sqrt{\Gamma_1 \Gamma_2}}{4}
  \left \{   
{\hat{\mathbf{\mu}}}_{1}\cdot  \hat{\mathbf{\mu}}_{2}
  -( \hat{\mathbf{\mu}}_{1}\cdot \hat{\mathbf{r}}_{12} )
  (\hat{\mathbf{\mu}}_{2}\cdot \hat{\mathbf{r}}_{12}) ] 
  \frac{\cos z}{z}  \right . \nonumber  \\ 
  & + & \left .  [ \hat{\mathbf{\mu}}_{1} \cdot  \hat{\mathbf{\mu}}_{2}
  -3(\hat{\mathbf{\mu}}_{1}\cdot \hat{\mathbf{r}}_{12})
  (\hat{\mathbf{\mu}}_{2}\cdot \hat{\mathbf{r}}_{12})]
  \left ( \frac{\cos z}{z^{3}}+\frac{\sin z}{z^{2}} \right )
\right \}, \nonumber \\
& & \\
  \gamma &  =&
  \frac{3\sqrt{\Gamma_1 \Gamma_2}}{2}
  \left \{ 
  [\hat{\mathbf{\mu}}_{1}\cdot  \hat{\mathbf{\mu}}_{2}
  -(\hat{\mathbf{\mu}}_{1}\cdot \hat{\mathbf{r}}_{12})
  (\hat{\mathbf{\mu}}_{2}\cdot  \hat{\mathbf{r}}_{12})] \frac{\sin z}{z}
\right .  \nonumber \\ 
  & + & \left . 
  [\hat{\mathbf{\mu}}_{1}\cdot  \hat{\mathbf{\mu}}_{2}
  -3 (\hat{\mathbf{\mu}}_{1}\cdot \hat{\mathbf{r}}_{12})
  (\hat{\mathbf{\mu}}_{2}\cdot \hat{\mathbf{r}}_{12})] 
  \left ( \frac{\cos z}{z^{2}}-\frac{\sin z}{z^{3}} \right ) \right \} 
  \, , \nonumber
\end{eqnarray}
where $z= nk_{0}r_{12}$,  $k_{0}= \omega_{0}/c$, 
and $\omega_{0}= (\omega_{1}+\omega_{2})/2$.

 We next consider that the emitters are located close to a plasmonic
waveguide. The new emission
properties, as well as their dipolar interaction strength, depend on
the coupling to the plasmonic modes.  The dipole-dipole
interaction strength and the incoherent decay rate can be calculated
from the Green's function $\mathrm{G}(\mathrm{r}_1,\mathrm{r}_2)$ that
describes the electromagnetic interaction between the two dipole
moments (emitters) $\mu_{1,2}$ of frequency
$\omega_{1}=\omega_{2}\equiv\omega_{0}$ as follows (see \cite{david}
for a discussion of the evaluation of these terms)
\begin{eqnarray}
  V^{\mathrm{pl}} & = & \frac{1}{\pi\epsilon_0 c^{2} \hbar}\mathcal{P}
  \int_0^{\infty}\frac{\omega^2\mathrm{Im}[\mu^\ast_1 
    \mathrm{G} (\omega,\mathbf{r}_1,\mathbf{r}_2)\mu_2]}
  {\omega - \omega_0 } d\omega \, , \nonumber  \\  
  \Gamma^{\mathrm{pl}}_{ij} & = &
  \frac{2\omega_0^2}{\epsilon_0c^2\hbar}\mathrm{Im}
  [\mu^\ast_i \mathrm{G} (\omega_0,\mathbf{r}_i,\mathbf{r}_j)\mu_j] ,
  \label{collective}
\end{eqnarray}
where the superscript ``pl'' indicates that the collective parameters
are now modified by the interaction with the plasmonic waveguide. As
before, we set $\Gamma^{\mathrm{pl}}_{ii}=\Gamma^{\mathrm{pl}}_i$, and
$\Gamma^{\mathrm{pl}}_{ij}=(\Gamma^{\mathrm{pl}}_{ji})^\ast =
\gamma^{\mathrm{pl}}$, with $i,j=1,2$. 

It has been recently shown~\cite{tudela,prb11} that both the dipolar interaction and
the collective damping for a pair of emitters close enough to a
plasmonic waveguide hold simple analytical forms when the dominant
contribution for emission is due to the plasmons. Thus, the Green
function can be approximated by the plasmonic Green function
contribution, $\mathrm{G} (\mathrm{r}_1,\mathrm{r}_2, \omega) \approx
\mathrm{G}^{\mathrm{pl}} (\mathrm{r}_1,\mathrm{r}_2,\omega)$. 

If we set the
plasmon wavelength $\lambda_{\mathrm{pl}}=2\pi/k_{\mathrm{pl}}$, we
have~\cite{david,tudela}
\begin{equation}
V^{\mathrm{pl}} =  \half \Gamma^{\mathrm{pl}}
\tilde{\beta}\sin{(2 \pi \zeta)} \, , 
\qquad
\gamma^{\mathrm{pl}}  =  \Gamma^{\mathrm{pl}} 
\tilde{\beta} \cos{(2 \pi\zeta)} \, ,
\end{equation}
 where $\tilde{\beta} = \beta \exp [-\lambda_{\mathrm{pl}} \zeta/(2 L)]$, with
$\zeta=d/\lambda_{\mathrm{pl}}$, $d$ is the distance between the
emitters, $\Gamma^{\mathrm{pl}}=\Gamma^{\mathrm{pl}}_{1}
=\Gamma^{\mathrm{pl}}_{2}$, $L$ is the propagation length of the
propagating mode, and the $\beta$ factor that measures the fraction of
emitted radiation by the propagating mode. It is worth mentioning
that, due to radiative contributions, such a `plasmonic approximation'
breaks down for emitters separations shorter than $\sim
\lambda_{\mathrm{pl}}/4$~\cite{david}.

We note from these two relations that $\mid V^{\mathrm{pl}}\mid \leq
\Gamma^{\mathrm{pl}}/2$ and
$\mid\gamma^{\mathrm{pl}}\mid\leq\Gamma^{\mathrm{pl}}$, which means that
the interaction that arises due to the coupling to the plasmons is
weak and the most important contribution due to collective effects
comes from the damping $\gamma^{\mathrm{pl}}$.

\subsection{Correlations}
\label{correla}

The definitions involved in the calculation of the correlations are as
follows.  The quantum mutual information describes the whole content
of correlations in a given quantum
system~\cite{popescu,hamieh,Vedral2}.  It has been shown that quantum
correlations (entanglement included)~\cite{Zurek} and classical
correlations~\cite{Vedral}, in the sense of entropic measures, add up
to give the quantum mutual information~\cite{hamieh}. Furthermore,
this point has been recently emphasized, via the use of the relative
entropy, within a unified framework that captures both quantum and
classical correlations within the quantum mutual
information~\cite{Vedral2}.  For a bipartite system, this can
be written as:
\begin{equation}
  I(\rho_{AB}) = S(\rho_A) +S(\rho_B)-S(\rho_{AB}) ,
  \label{MI}
\end{equation}
where $S(\rho)=-\Tr ( \rho \log_{2} \rho)$ is the von Neumann entropy
of density matrix $\rho$.

A measure of classical correlations was introduced in Ref.~\cite{Vedral} as
the maximum extractable classical information from a subsystem, say
$A$, when a set of positive operator valued measures~\cite{man} has
been performed on the other subsystem (say $B$):
\begin{equation}
  \label{CC}
  \mathrm{CC}(\rho_{AB}) = \sup_{\{\Pi_j^B\}} \left [
    S(\rho_{A})-\sum_j p_j S(\rho_{A}^j)\right ]  ,
\end{equation}
where $S(\rho_{A}^j)$ is the entropy associated to the density matrix
of subsystem $A$ after the measure. Such correlations must be
non-increasing, and invariant under local unitary operations, and
$\mathrm{CC}(\rho_{AB}) = 0$ if and only if $\rho_{AB}= \rho_A\otimes
\rho_B$.  

If $\{\ket{0},\ket{1}\}$ define the basis states for the qubit 
$B$, the projectors can be written as $\Pi_j^B=\mathrm{\bf 1}\otimes
\proj{j}$, $j=a,b$, where $\ket{a}= \cos{\theta}\ket{0}+e^{\rm
  i\phi}\sin{\theta}\ket{1}$, $\ket{b}= e^{- i\phi}
\sin{\theta}\ket{0}-\cos{\theta}\ket{1}$, and the optimization is
carried out over angles $\theta$ and $\phi$. The measure $\mathrm{CC}$
is antisymmetric by definition, and, without loss of generality, we
take the qubit $B$ to be the one measured.

Following the definition for $\mathrm{CC}(\rho_{AB})$, a simple way to
define the total quantum correlations in a bipartite system is
$D(\rho_{AB})=I(\rho_{AB})-\mathrm{CC}(\rho_{AB})$. In terms of 
the von Neumann entropies, the quantum correlations, which coincide
with the definition for the quantum discord given in Ref.~\cite{Zurek},
read
\begin{equation}
  \label{D}
  D(\rho_{AB})= S(\rho_B)-S(\rho_{AB}) +
  \inf_{\{\Pi_j^B\}} \sum_jp_j S(\rho_{A|\Pi_j^B}) \, .
\end{equation}
For pure states, $D=S(\rho_B)$, and $D=0$ iff the system is purely
classically correlated.

Quantum entanglement is quantified by the entanglement of formation
\begin{equation}
  \mathrm{EoF} (\rho)=h \left (\frac{1 +\sqrt{1 - C^2(\rho)}}{2}
  \right) ,
  \label{EoFR}
\end{equation}
where $ h(x)=-x\log_2x-(1-x)\log_2(1-x) $ denotes the binary entropy
function~\cite{wootter}. 

Additionally, consider, in decreasing order, the
eigenvalues $\lambda_i$ of the matrix $\sqrt{\rho_{AB}
  \tilde\rho_{AB}}$, where $\tilde\rho_{AB} =
(\sigma_y\otimes\sigma_y) \bar\rho_{AB} (\sigma_y\otimes\sigma_y)$ and
$\bar\rho_{AB}$ is the elementwise complex conjugate of $\rho$. The
concurrence $C$ can be defined as
\begin{equation}
  C(\rho_{AB})={\mathrm{max}}\{0, \lambda_1- \lambda_2- \lambda_3-\lambda_4\} , 
  \label{concurrence}
\end{equation}
where the $\lambda_i$'s are as introduced above or, equivalently (also
in decreasing order), the square root of the eigenvalues of the
non-Hermitian matrix $\rho_{AB} \tilde\rho_{AB}$~\cite{wootter}.

\section{Dynamics of correlations}

We calculate the exact dynamics of correlations by solving the master
equation~(\ref{master}), and taking into account the definitions
introduced in section~\ref{correla}.  The notation for the
correlations reported in the graphs throughout this paper is as
follows: total correlation or mutual information (purple, thick-solid
line), classical correlation (green, doubly-dashed line), quantum
discord (blue, thin-solid line), entanglement of formation (red,
dashed line), and concurrence (pink, dotted line).

To compare  the physical scenarios described in
Sec.~\ref{system}, we first present correlations when the emitters
interact solely with the vacuum electromagnetic field, i.e., in the
absence of coupling to plasmons. In terms of the collective effects,
such a dynamics depends on the separation between emitters and also on
the orientation of their dipole moments.

\begin{figure}
  \centering
  \includegraphics[width=0.60\columnwidth]{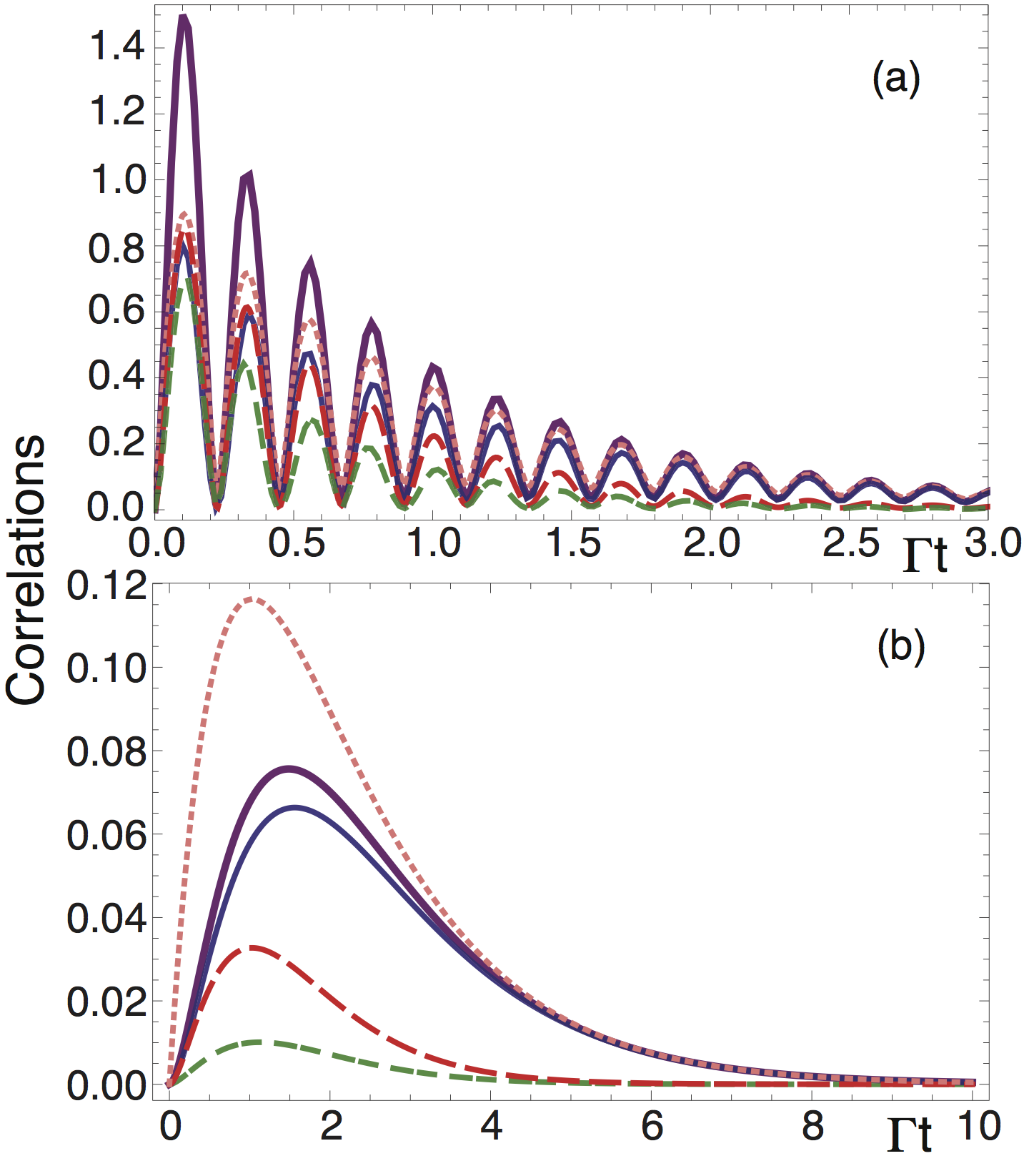}
  \caption{(Color online) Correlations for a two-emitter system in the
    {\it absence} of coupling to plasmons.  (a) $V=7 \Gamma$ and
    $\gamma = 0.2 \Gamma$. (b) Dipoles parallel to each other and
    perpendicular to their separation vector;
    $r_{12}=\frac{3}{4}\lambda_0$, $\lambda_0\equiv 2\pi/k_0$. The
    considered initial state is separable: $\rho(0) =
    \ket{10}\bra{10}$. In all graphs presented in this work: Total
    correlation or mutual information (purple, thick-solid line),
    classical correlation (green, double-dashed line), quantum discord
    (blue, thin-solid line), entanglement of formation (red, dashed
    line), and concurrence (pink, dotted line). }
  \label{fig1}
\end{figure}

A typical situation where the dipole-dipole interaction is much
greater than the incoherent part $V \gg \Gamma \gg \gamma$
[figure~\ref{fig1}(a)] can be observed, for example, in a system of
diluted molecules of Terrilene in a dispersive crystal~\cite{Hettich},
for which a rich entanglement dynamics has been
  predicted~\cite{jh3}.  Figure~\ref{fig1}(a) shows that correlations
reach high values and oscillate rapidly because of the strong
dipole-dipole interaction energy $V$; however, the correlations decay
within a short time and the system becomes uncorrelated quite rapidly.

A different scenario, depicted in figure~\ref{fig1}(b), considers
a pair of dipoles parallel to each other and perpendicular to their
separation vector, which allows the calculation of the collective
parameters $V$ and $\gamma$ directly from Eqs.~(\ref{V}). Although the
correlations initially show smaller values than before, they persist
for a longer time.

An interesting issue that arises from figure~\ref{fig1} is the
discrepancy between the entanglement of formation and the concurrence
as quantifiers of entanglement.  Figure~\ref{fig1}(b) shows that
concurrence reaches much higher values (and decays much slower) than
the EoF; more surprising is the fact that concurrence reaches, for
almost any time, higher values than the total correlations, a result
that contrasts with the definition of the mutual information as a
quantity that accounts for all the correlations (classical and
quantum), entanglement included~\cite{hamieh,Vedral2}. Thus, within
this framework, concurrence can indicate results that are well above
the EoF, and hence does not allow a direct comparison with other
entropic measures such as the quantum discord; in contrast, the latter
can be compared, on the same grounds, to the EoF
\cite{fanchini2}. This is explicitly shown in figure~\ref{fig1}(b).

\begin{figure}[t]
  \centering
  \includegraphics[width=0.60\columnwidth]{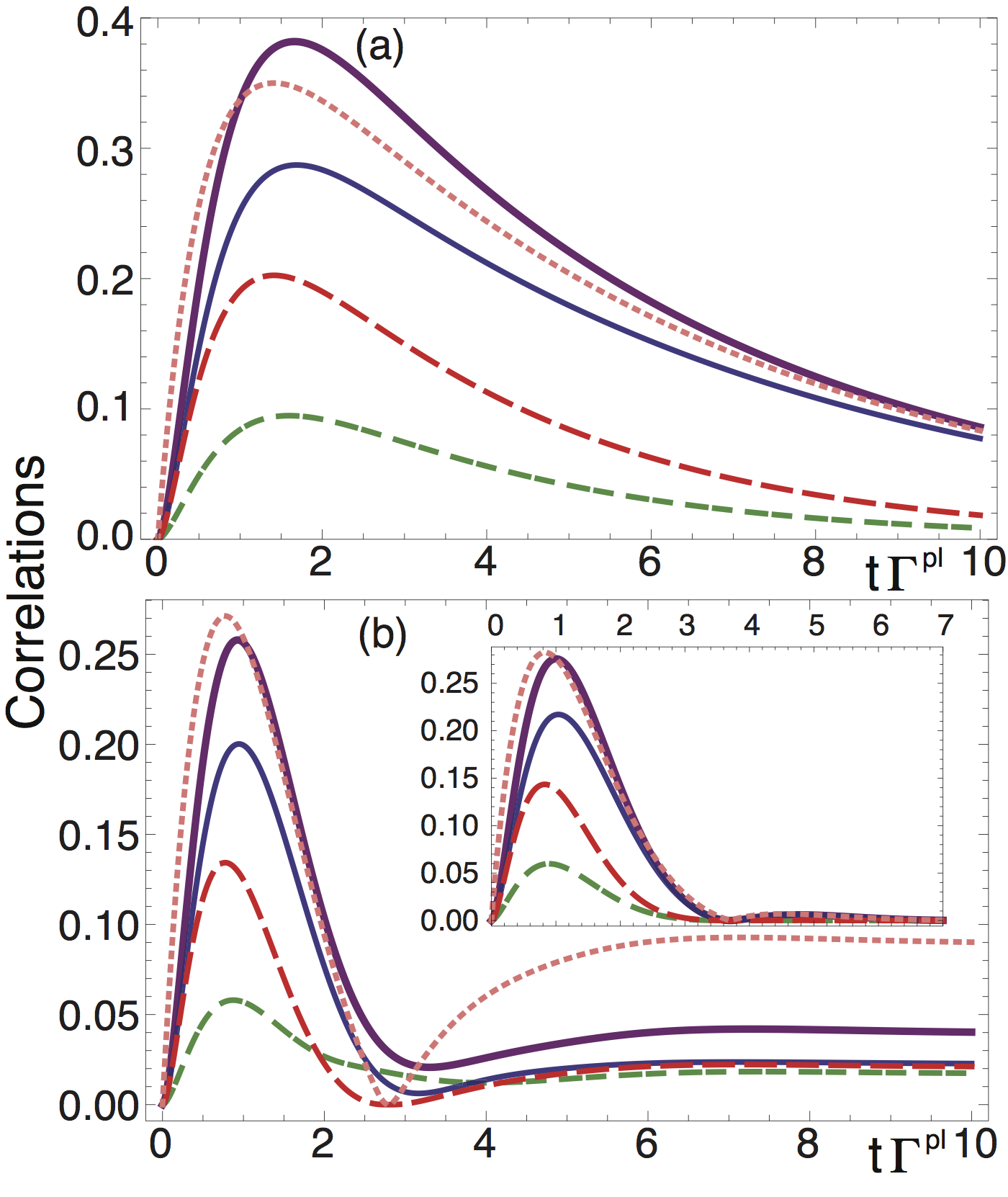}
  \caption{(Color online) Correlations for the emitters interacting
    with a plasmonic waveguide with the experimentally feasible set of
    parameters $\beta=0.94$, $L=2\mu$m, obtained for an operational
    wavelength $\lambda_0=640$ nm; this gives $\tilde{\beta}
    \simeq0.82 $, and a plasmonic wavelength $\lambda_{\mathrm{pl}}
    \simeq 542$ nm. (a) Separation between emitters
    $d=\lambda_{\mathrm{pl}}$, $V^{\mathrm{pl}} =0$, (b) $\zeta\equiv
    d/\lambda_{\mathrm{pl}} =3/4$, $\gamma^{\mathrm{pl}} =0$, and
    $\ell=0.2 \, \Gamma^{\mathrm{pl}}$.  The inset corresponds to case
    (b), but in the absence of laser pumping. The notations for the
    graphs are as given in figure~\ref{fig1}.}
  \label{fig2}
\end{figure}

The influence of the
plasmonic waveguide is illustrated in figure~\ref{fig2}, and can be
directly compared with the results in figure~\ref{fig1}.
According to the relations derived from
Eqs.~(\ref{collective})~\cite{david,tudela}, one of the parameters
$V^{\mathrm{pl}} $ or $\gamma^{\mathrm{pl}} $ can, in principle, be
maximized by `switching off' the other one. For the interqubit
distance $d = \frac{3}{4} \lambda_{\mathrm{pl}} $ [Fig \ref{fig2}(b)],
the plasmonic modes allow for an effective enhancement of correlations,
as can be seen from a direct comparison between the inset of
figure~\ref{fig2}(b) and figure~\ref{fig1}(b). This said, note that the
correlations decay more rapidly in the plasmon-assisted case because
here the emitters distance is such that the collective rate is
switched off, and the nonlocal effects are purely due to the weak
dipole-dipole interaction $V^{\mathrm{pl}} $. The correlations are enhanced as
$\tilde{\beta}$ tends to $1$. For an emitters separation
$d=\lambda_{\mathrm{pl}} $, the realistic value $\tilde{\beta} = 0.82$
can be obtained for $L=2\, \mu$m, and this is shown in
Fig. \ref{fig2}(a): for this interqubit distance, $V^{\mathrm{pl}} $
is switched off and the correlations reach larger values than before
and are maintained for a much longer time.  We also point out that for
the parameter window considered in figure~\ref{fig2}, classical
correlations do not play a major role.

Interestingly, the quantum discord, which is in its essence different
to entanglement, does play a role (and is more robust than the EoF)
during the qubit dissipative evolution.  We remark that is not
  only the existence of quantum correlations, but the way they relate
  to each other, that matters, since the latter can lead to
operational interpretations~\cite{fanchini2}.  In fact,
figure~\ref{fig2}(a) shows that for a time $\tau=10/\Gamma^{\mathrm{pl}}
$, entanglement, as calculated by the EoF, has almost vanished and the
quantum discord quickly approximates to the value given by the total
correlations. In contrast to this, Ref.~\cite{tudela} uses the
concurrence to show that a large amount of entanglement is generated
for this set of parameters. This can be seen in figure~\ref{fig2}(a),
where it is evident that the concurrence approximately equals
the total correlations, in clear contrast to what is shown by the EoF.
Within the formal framework of correlations presented here, the
emitters entanglement is well below that reported in
  Ref.~\cite{tudela}.   This is consistent with the approach of entropic
measures followed here. 

Furthermore, it is straightforward to show that the density matrix
computed at $t=\tau$ (figure~\ref{fig2}(a)) corresponds to the mixed
state $\rho(\tau)=0.9166 \ket{00}\bra{00} +
0.0834 \ket{\Psi^-}\bra{\Psi^-}$, where 
$\ket{\Psi^-}= \left(\ket{01} - \ket{10}\right)/\sqrt{2}$ is
the subradiant state. This state has a very small contribution 
from the entangled state $\ket{\Psi^-}$, and so it provides a small
degree of entanglement. However, the mixture of this with the product
state $\ket{0}\otimes \ket{0}$, produces a quantum correlated state
with a type of correlation---the so-called quantum discord, which is
different, in its very essence, to entanglement~\cite{Zurek,Vedral2}. 

Although a larger value of $\tilde{\beta}$ can enhance entanglement as
well as the other correlations (not shown), we point out that care
must be taken when choosing such a parameter; in particular, the value
$\tilde{\beta}=0.9$ used in \cite{tudela} is not realistically
attainable according to the plasmon dispersion relation reported there
for $d=\lambda_{\mathrm{pl}} $.  Indeed, if we include the
  `best' achievable $\beta(\simeq 0.94)$ factor, in the expression for
  $\tilde{\beta}\rightarrow0.9$, the relation between the plasmonic
  wavelength and the propagation length should read
  $\lambda_{\mathrm{pl}} /L\simeq0.08697$, but from the known
  dispersion relation~\cite{tudela}, no pair $\{\lambda_{\mathrm{pl}}
  \, ,\, L\}$ satisfies this constraint. This simple estimation leads
us to conclude that, if $\beta \sim 0.94$, the largest correct
realistic value that should be used for the considered emitters
separation is $\tilde{\beta}\simeq 0.82$.

\begin{figure}
  \centering
 \includegraphics[width=0.6\columnwidth]{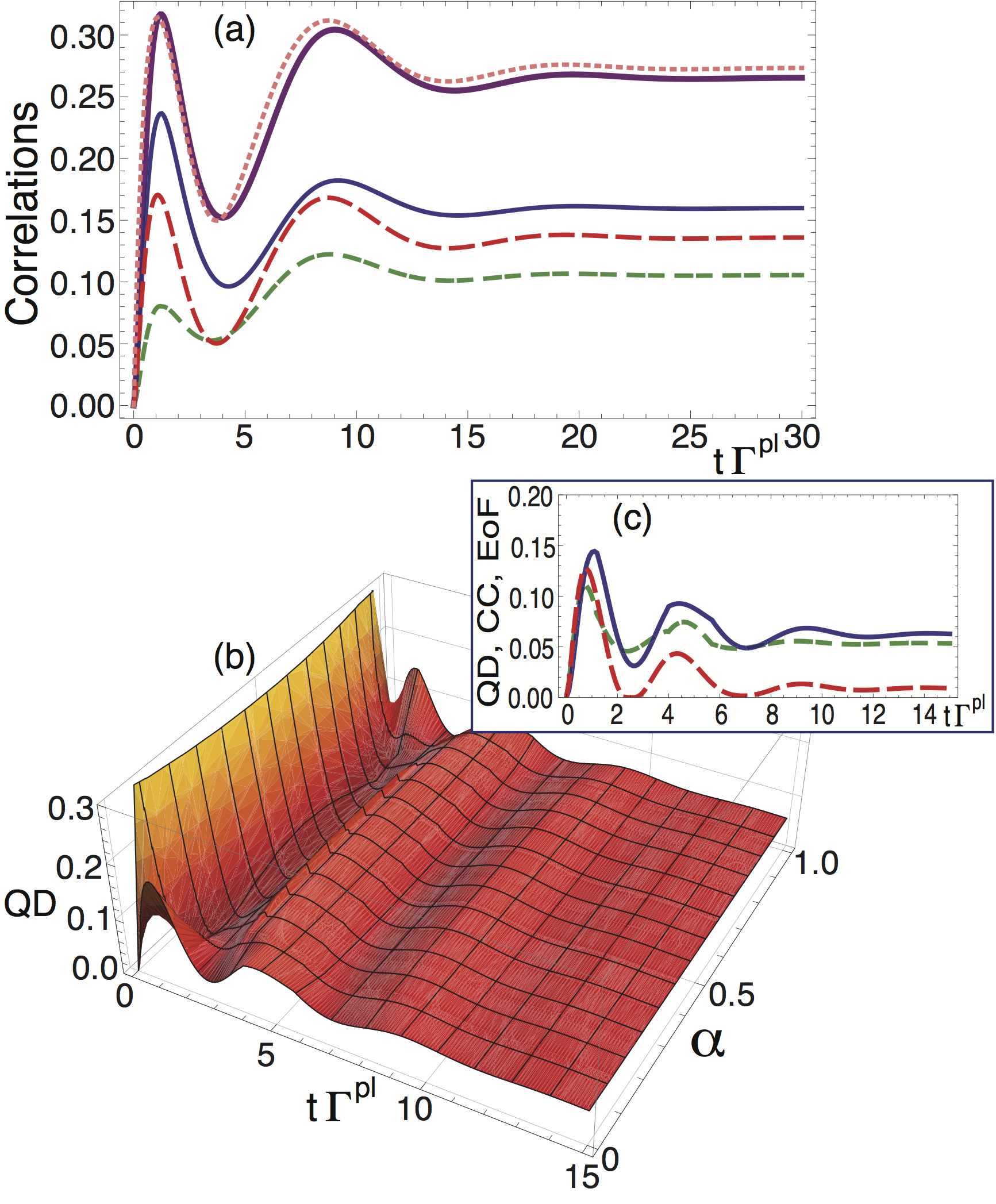}
  \caption{Optical control of correlations dynamics. (a) Total
    spectrum of correlations for laser amplitude
    $\ell_{1}=-\ell_{2}=\ell=0.2 \, \Gamma^{\mathrm{pl}} $, and
    initial state $\ket{01}$. (b) Quantum discord dynamics for
    $\ell_{1}=-\ell_{2}=\ell=0.4 \, \Gamma^{\mathrm{pl}} $, and
    initial state configurations $\sqrt{\alpha}\ket{01} +
    \sqrt{1-\alpha}\ket{10}$. (c) Comparison between QD, CC, and EoF
    for $\alpha = 1$.  In all graphs, $\zeta=1$, $V^{\mathrm{pl}} =0$,
    $\beta=0.94$, and $L=2\mu$m.}
  \label{fig3}
\end{figure}

The qubit correlation dynamics due to the collective effects arising
from the emitters coupling to the plasmon modes, $V^{\mathrm{pl}} $
and $\gamma^{\mathrm{pl}} $, become even more interesting when they
are pumped with a continuous laser field of amplitude $\ell_i$ and
frequency $\omega_i$ (targetting the emitter $i$). If we consider the
laser excitation to be in resonance with the emitters  transition
frequency $\omega_i=\omega_0$, stationary correlations can be obtained
by making the amplitudes $\ell_1=\ell_2\equiv\ell$, as shown in
figure~\ref{fig2}(b) for a distance $\zeta = 3/4$, $\gamma^{\mathrm{pl}}
= 0$.  In the same spirit, a higher stationary behaviour of
correlations is obtained for $\zeta = 1$, by introducing a relative
phase between the laser amplitudes: $\ell_1 = -\ell_2 = \ell$, as
shown in figure~\ref{fig3}(a). The correlations are much larger than
those obtained in figure~\ref{fig2}(b) because the laser excitation
assists the very slow decay of the `naturally created' antisymmetric
state $\ket{\Psi^-}=(\ket{01}-\ket{10})/\sqrt{2}$. This
figure~also shows that the concurrence reaches higher values than the
total correlations, which might be seen as inappropriate when compared
to the use of entropic metrics.

We have shown that rather than entanglement, quantum discord is the
most robust correlation during the dissipative dynamics,
and, depending on the separation between emitters, the qubit-plasmon
modes coupling can enhance the degree of correlations via the
collective parameter $\gamma^{\mathrm{pl}} $. For the sake of completeness, in
figure~\ref{fig3}(b) we consider, for a fixed laser amplitude, the
role of the initial state preparation on the correlations dynamics.
Here, we plot the quantum discord as a function of time, for initial
states $\sqrt{\alpha}\ket{01}+\sqrt{1-\alpha}\ket{10}$, $\alpha \in
\mathbb{R}$.  Although it is clear that, at $t=0$, the discord takes the
value 1 for $\alpha=1/2$, we have plotted it only for a range up to
$D=0.3$, in order to appreciate comparatively the correlations
behaviour for all $\alpha$. The trend in correlations is clearly
influenced by $\alpha$ but, overall, the states converge to a common
stationary value. The quantum discord, classical correlations, and EoF
are plotted for the particular case $\alpha=1$ in
figure~\ref{fig3}(c), showing how an increment in the laser amplitude
may produce a noticeable difference between the quantum discord and
entanglement (EoF). Thus, pumping with a continuous laser field can be
used as an additional means to dynamically control the relationship
between entanglement and the quantum correlations present in the
emitters. Such a fact could be used as an important resource in the
realisation of non-conventional quantum protocols~\cite{dqc1,datta,zu}.

Bearing in mind that the quantum correlation, as measured by the
discord, gives the amount of information that is not accessible to
local measurement, our results show that most of the information
stored in the emitters is purely quantum (compare the thin-solid blue
with the doubly-dashed green lines in all the figures). Also, we note
that, in most cases, this type of information does not arise from
entanglement, which is kept to a minimum, as can be seen in
Fig.s~\ref{fig1}(b), \ref{fig2}(a), and \ref{fig3}(c). This said, a
possible optical control of the amount and class of induced quantum
correlations can be carried out with external laser pumping, as
depicted in figure~\ref{fig3}.

\section{Discussion}

A comparison of the two physical scenarios considered here ---emitters
dissipative dynamics in the i) absence and ii) presence of coupling to
a plasmonic bus--- leads us to the following. In the former case, the
correlations exhibit a longer lifetime in figure~\ref{fig1}(b) because
of the presence of the incoherent interaction $\gamma$ for the
distance $r_{12}=\frac{3}{4} \lambda_0$; this means that the lifetime
of the symmetric correlated state $\ket{\Psi^+}=
(\ket{01}+\ket{10})/\sqrt{2}$ (an eigenstate of the bare
Hamiltonian ${H}_S$) is more robust ($\gamma<0$). In the latter
scenario, $\gamma^{\mathrm{pl}} = 0$ and the correlations are only
mediated by the coherent interaction $V^{\mathrm{pl}} $. The
plasmon-mediated coupled emitters become more interesting because for
large distances ($d\sim\lambda_{\mathrm{pl}} $), the incoherent decay
$\gamma^{\mathrm{pl}} $ can take values close to $\Gamma^{\mathrm{pl}}
$, holding the correlations for longer times. A particular case is
shown in figure~\ref{fig2}(a): $d = \lambda_{\mathrm{pl}} $,
$\gamma^{\mathrm{pl}} =0.82 \Gamma^{\mathrm{pl}} $, hence
$V^{\mathrm{pl}} $ is `switched off'. The correlations are present
even for times of the order of $\Gamma^{\mathrm{pl}} t\sim10$ thanks
to the slow decay $\Gamma^{\mathrm{pl}} -\gamma^{\mathrm{pl}} $ of the
antisymmetric state $\ket{\Phi^-}$. For this time scale, entanglement
(as measured by the EoF) is almost zero, and the correlation that
prevails is the quantum discord, which in turn tends to the same value
as the total correlations.

Aside from the physics that arises in the qubit-plasmon setup here
considered, we highlight the discrepancy between the EoF and the
concurrence as quantifiers of  the emitters entanglement.  This
is due to the entropic origin of the EoF compared to that of
concurrence (the difference between the two is explicitly stated in
equation~\ref{EoFR}).  Furthermore, in most of the plots of this
  work, the concurrence is larger than the mutual information (purple
lines) in several different time frames. Since the quantum mutual
information captures all the possible correlations (entanglement
included), this result seems incorrect. Again, the explanation for
this lies in the fact that the concurrence does not have an entropic
origin like the other correlation measurements calculated here.
Strictly speaking, the concurrence is an entanglement monotone but is
not an `actual' entanglement metric in the sense that it obtains its
meaning from its relation to the entanglement of formation and not the
opposite \cite{plenio2007}. With the emergence of quantum correlations
beyond the entanglement, it is crucial to coherently quantify the
latter in order to give a consistent physical interpretation of
quantum phenomena.  The understanding of whether a qubit system is
entangled or discord-correlated is thus relevant for deciding the way
in which the emitters could be operated as a physical device for
performing information processing tasks.

The relation between the correlations and the physical properties of
the hybrid system analyzed here is clearly showed in
Figs.~\ref{fig1}-\ref{fig3}: the lifetime of the correlations is
enhanced thanks to the illumination with the coherent laser field;
also, a large $\beta$-factor (due to the plasmon channel) increases
the degree of correlations, especially that of the quantum discord. It
is evident from figure~\ref{fig3} that an adequate manipulation of the
light-matter interaction strength allows control of the dynamics
followed by the quantum discord and entanglement, and therefore of
their metric value.  This degree of quantum control is possible due to
the interplay between the laser illumination intensity and the
coherent emitter-emitter interaction \cite{jh3,jh2011}.

\section{Concluding remarks}

We have computed entanglement, classical, quantum, and total
correlations for a hybrid system composed of  largely separated emitters coupled to
  the plasmonic modes of a one-dimensional waveguide, which is
externally driven by a laser field.  We have illustrated, by direct
calculation, that classical correlations are the least assisted by the
plasmon bus and that quantum discord is the dominant
correlation that prevails throughout the whole dissipative dynamics
and that this is enhanced by the presence of the plasmonic collective
excitations and by the external laser pumping. This tendency in the
correlations has been analytically demonstrated for bare emitters
coupled to the vacuum electromagnetic field in
  Ref.~\cite{jh2011}.

We have emphasized the entropic  origin of the entanglement of
formation, the quantum discord, and that of the quantum mutual
information as quantifiers of correlations; we also computed the
concurrence for direct comparison of our results with those reported
in \cite{tudela,prb11}. We found that the latter can reach values well
above the EoF and the total correlations. This shows that care must be
taken when confronting or using specific quantifiers of entanglement
for describing a physical process, especially because the
interpretation of the EoF as the cost of creation of an entangled
state (with no regularization), leads to an upper bound on the actual
degree of entanglement in a quantum system~\cite{cornelio}.

We remark that a crucial feature of the mechanism devised  here is the
large separation between emitters (of the order of or greater than the
plasmons wavelength), which could lead to long-distance quantum
control and communication: the correlations are purely generated via
the strong plasmon-emitters interaction, which in turn is reflected in
the functional dependence of the collective parameters given by
equation~(\ref{collective}). An interesting by-product is the
feasibility of tailoring the phase difference that exists between the
collective parameters, which might allow the switching on/off of the
emitters correlations. Here, a parameter configuration where either
$V^{\mathrm{pl}}$ or $\gamma^{\mathrm{pl}}$ goes to zero, an appropriate choice of initial
conditions and laser tailoring, enables the enhancement or suppression
of the existing correlations. This result is beyond the scope of this
work and will be addressed elsewhere \cite{qswitch12}.

Recent experiments have demonstrated the control of light-matter
interaction by means of plasmonic resonators \cite{leon} and silver
nanowires \cite{kumar} to enable the enhancement of the Purcell effect
in quantum emitters, and the strength of the coupling to the plasmonic
modes of such nanostructures. This offers the possibility of testing
the results here reported with state-of-the-art
technology. Furthermore, another experimental demonstration of
communication between two distant single emitters (organic molecules)
via single photons has been recently reported \cite{rezus}.  Thus, the
setup proposed in this work has also the potential for demonstrating
an additional degree of quantum control in which sensitive quantum
information encoded in single photons can be transmitted and processed
between coupled (but largely separated) single emitters.

\ack

 C.E.S. is grateful for a Colciencias fellowship. We acknowledge partial  financial support from Universidad del Valle
(Grant CI 7859), the Spanish DGI
(Grant FIS2011-26786), and the UCM-BSCH program
(Grant GR-920992).

\end{document}